# Long-lived populations of momentum- and spin-indirect excitons in monolayer WSe$_2$

*Shao-Yu Chen,[1,2] Maciej Pieczarka,[1,3,4] Matthias Wurdack,[1,3] Eliezer Estrecho,[1,3] Takashi Taniguchi,[5] Kenji Watanabe,[6] Jun Yan,[7] Elena A. Ostrovskaya,[1,3] and Michael S. Fuhrer[1,2,*]*

[1]ARC Centre of Excellence in Future Low-Energy Electronics Technologies

[2]School of Physics and Astronomy, Monash University, Clayton, Victoria, 3800, Australia

[3]Nonlinear Physics Centre, Research School of Physics, The Australian National University, Canberra, ACT 2601, Australia

[4]Department of Experimental Physics, Wrocław University of Science and Technology, Wyb. Wyspiańskiego 27, 50-370 Wrocław, Poland

[5]International Center for Materials Nanoarchitectonics, National Institute of Materials Science, 1-1 Namiki, Tsukuba, Ibaraki 305-0044, Japan

[6]Research Center for Functional Materials, National Institute of Materials Science, 1-1 Namiki, Tsukuba, Ibaraki 305-0044, Japan

[7]Department of Physics, University of Massachusetts, Amherst, MA 01003, USA

[*]Corresponding Author: Michael S. Fuhrer

Tel: +61 3 9905 1353

E-mail: Michael.Fuhrer@monash.edu

**Abstract**

Monolayer transition metal dichalcogenides are a promising platform to investigate many-body interactions of excitonic complexes. In monolayer tungsten diselenide, the ground-state exciton is dark (spin-indirect), and the valley degeneracy allows low-energy dark momentum-indirect excitons to form. Interactions between the dark exciton species and the optically accessible bright exciton (X) are likely to play significant roles in determining the optical properties of X at high power, as well as limiting the ultimate exciton densities that can be achieved, yet so far little is known about these interactions. Here, we demonstrate long-lived dense populations of momentum-indirect intervalley ($X_K$) and spin-indirect intravalley (D) dark excitons by time-resolved photoluminescence measurements (Tr-PL). Our results uncover an efficient inter-state conversion between X to D excitons through the spin-flip process and the one between D and $X_K$ excitons mediated by the exchange interaction (D + D $\leftrightarrow$ $X_K$ + $X_K$). Moreover, we observe a persistent redshift of the X exciton due to strong excitonic screening by $X_K$ exciton with a response time in the timescale of sub-ns, revealing a non-trivial inter-state exciton-exciton interaction. Our results provide a new insight into the interaction between bright and dark excitons, and point to a possibility to employ dark excitons for investigating exciton condensation and the valleytronics.

**Keywords**

Excitons in two-dimensional (2D) hexagonal transition metal dichalcogenides (h-TMDs) feature large binding energy and strong light-matter interaction, while strong spin-orbit coupling and the two-fold valley degeneracy lead to optical, spintronic and valleytronic properties.[1,2] Excitons in monolayer h-TMDs have orders of magnitude smaller effective mass compared with *e.g.* alkali atoms, and have been predicted to achieve Bose-Einstein condensation (BEC) at critical temperatures up to 80−100 K.[3,4] However achieving the high exciton densities and strong exciton-exciton interactions necessary for condensation is challenging. Owing to the giant oscillator strength of the bright exciton in monolayer h-TMD, the ultrashort population lifetime of less than 2 ps leads to a significant drop of exciton density before thermalization takes place.[5] One approach to overcoming this issue is to implement a spatially indirect excitonic system in double-layer h-TMD heterostructures.[3,4,6,7] Having the electrons and holes in the different h-TMD layers, the spatially-indirect exciton exhibits much longer radiative lifetime of the order of ns.[8] Recently, the signatures of the exciton BEC in such a system were reported through the observation of enhanced tunnelling conductance at 100 K.[9] In the tungsten-based h-TMDs, in contrast to their molybdenum counterparts, the lower-energy dark excitons have 2−3 orders longer population lifetime.[10] It is therefore of interest to explore the possibility of achieving high density, strongly interacting dark exciton populations in the tungsten-based h-TMDs, which could host exciton BECs.[7] Long-lived dark excitons could also be useful as robust carriers of spin or valley information in spintronic or valleytronic devices through manipulating the helicity of the emitting photons.[11,12] The first step towards these applications is to understand the dynamics of dark excitons and the exciton-exciton interactions in tungsten-based h-TMDs.

In monolayer tungsten diselenide (1L-WSe$_2$), the strong spin-orbital coupling of d-orbitals of W atoms results in the energy splitting of 38 meV and 460 meV in the conduction and valence band, respectively.[13,14] The spin and valley configurations of various type of excitons in 1L-WSe$_2$ are illustrated in Figure 1a, and the corresponding quasiparticle band structures are shown in Figure 1b. To simplify the discussion, we only consider the top valence band and the case of holes in K valley; the exciton with holes in −K valley can be derived by the corresponding time-reversal pairs. The intravalley exciton X (as marked by the solid pink line in Figure 1a), also known as the bright exciton, has a total spin $S = 0$

(spin ½ electron plus spin -½ hole). As can be seen in Figure 1b, X exciton in 1L-WSe₂ is not the two-particle ground state. Upon photon excitation, the hot X excitons thermalise to other energetically favourable states. The spin-forbidden intravalley D exciton ($S = 1$, marked by the dashed green line) has an energy about 40 meV below the X exciton.[15] The D exciton has much weaker but finite oscillator strength with the out-of-plane dipoles and thus can still be accessed optically.[10,16] Several studies have revealed the interesting optical and valleytronic properties of the D exciton.[17,18] The intervalley $X_K$ exciton ($S = 0$, marked by the dotted blue line) is composed of electrons and holes with the same energy as the D exciton. However, the exchange interaction raises the binding energy of $X_K$ about 10 meV,[19,20] yielding the estimated energy of 30 meV below X exciton. $X_K$ exciton has the centre of mass momentum of K. To be radiatively recombined, it requires the coupling to zone-boundary phonons with momentum K (~31 meV)[21] to scatter the exciton into the light cone (the blue zigzag arrow in Figure 1b), further reducing the oscillator strength. As a result, the energy of a photon emitted by the $X_K$ exciton is 61 meV below the X exciton emission, even lower than D exciton.[21–24]

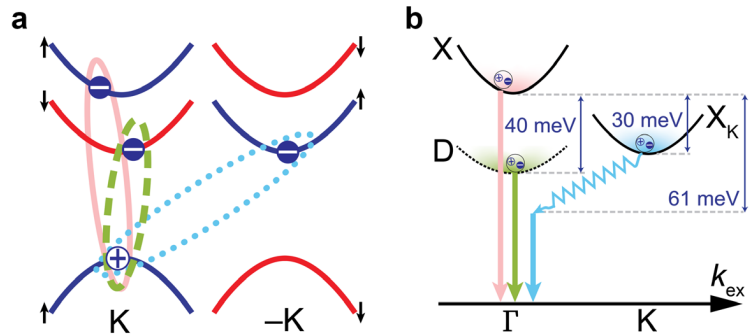

**Figure 1. Three types of charge-neutral excitons in 1L-WSe₂. a**, the spin-valley configuration of the intravalley X ($S = 0$, pink solid), intravalley D ($S = 1$, green dashed), and intervalley $X_K$ ($S = 0$, blue dotted) excitons. **b**, the schematic of excitonic band structure of X, D, and $X_K$ excitons. X and D are the intravalley excitons at $k_{ex} = \Gamma$ while $X_K$ is the intervalley exciton at $k_{ex} = $ K. The energy of $X_K$ is estimated around 30 meV below X but the emission photon energy is measured 61 meV below X due to the extra phonon energy about 31 meV.

In this work, by measuring the time-resolved photoluminescence (Tr-PL) of the high-quality hexagonal boron nitride (hBN) encapsulated 1L-WSe$_2$ at cryogenic temperature, we observe dense populations of D and X$_K$ excitons with sub-ns lifetime. Notably, we find that the X$_K$ exciton exhibits much slower growth rate and superlinear fluence-dependent dynamics. These observations indicate that X$_K$ exciton formation and lifetime are strongly governed by a second-order exciton-exciton interaction (D + D $\leftrightarrow$ X$_K$ + X$_K$) *via* Coulomb exchange. Moreover, as the exciton density increases, we observe a redshift of the X exciton sustained up to sub-ns, distinct from the ultrafast response reported before.[25–27] Most interestingly, the magnitude of redshift is proportional to the square of the excitation density, which is highly consistent with the population density of the X$_K$ exciton. Assisted by the rate equation analysis, we argue that the long-lived redshift of X exciton is caused by the excitonic screening effect, mainly due to the X$_K$ excitons, reflecting the strong inter-state exciton-exciton interaction as well as the dark exciton-mediated Coulomb screening. Our findings reveal the capability to create a long-lived, high-density population of momentum- and spin-indirect dark excitons for studies of excitonic many-body physics and exciton BEC.

**Results and Discussion**

The hBN-encapsulated 1L-WSe$_2$ samples are made by the polymer-based dry transfer technique in a nitrogen-filled glovebox (see Methods for more detail). After stacking, the samples are further thermally annealed in the argon atmosphere at 350 °C for 1 hour to remove the polymer residue. The sample is then transferred to a continuous flow cryostat with optical access and cooled down with liquid helium to the base temperature of 4.2 K. The details of the experimental setup of Tr-PL are described in Methods. Briefly, we excite the sample with a linear-polarised pulsed laser with a pulse width of 140 fs at various photon energies. The collected PL signal is filtered by a thin-film long-pass filter and then dispersed spectrally by a monochromator. The signal is detected by a thermoelectric-cooled charge-coupled device (CCD) for measuring the spectra. For measuring Tr-PL, the signal is redirected to a streak camera for acquiring the evolution of the PL emission in both time and frequency domains.

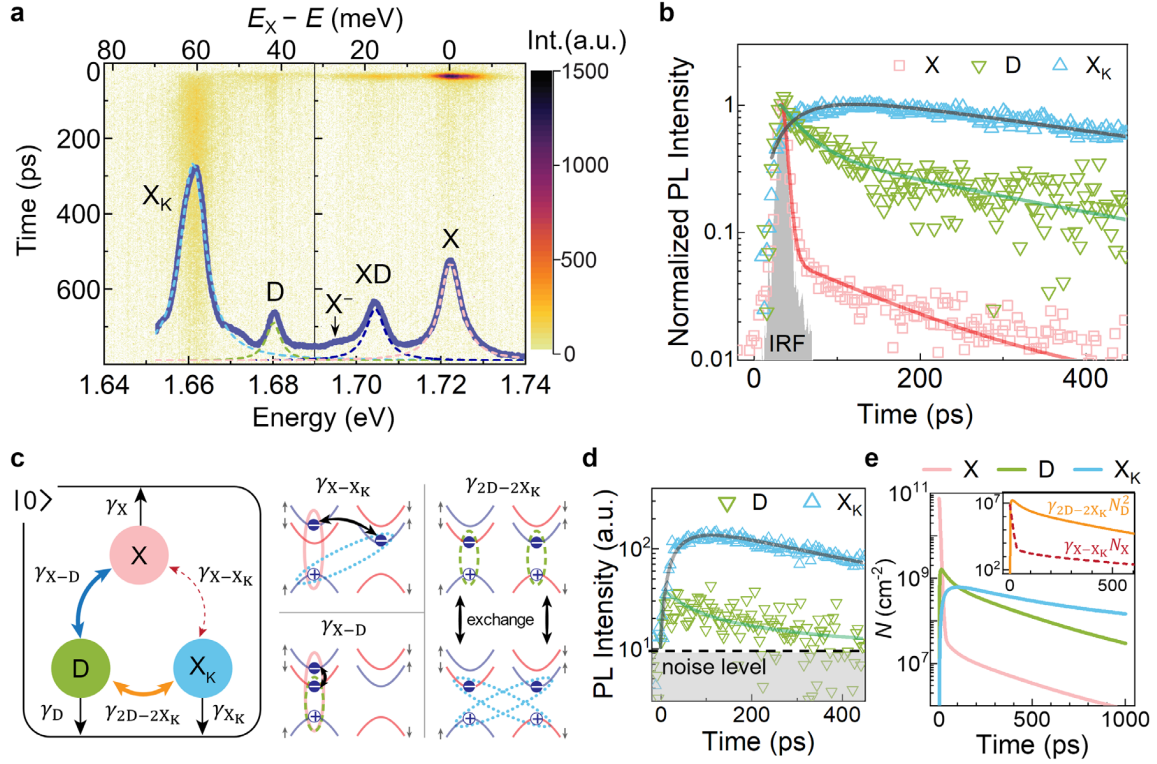

**Figure 2. Time-resolved photoluminescence of X, D, and $X_K$ excitons. a**, the contour plot of Tr-PL spectra at pump fluence of 3.4 μJ/cm⁻² at 2.75 eV. The associated PL spectrum is overlaid for comparing the integrated PL intensity. **b**, the normalized PL intensity of X, D, and $X_K$ excitons are fitted by double exponential decay. **c**, left: The schematic of four-level (X, D, $X_K$, and ground state) rate equation model for describing the exciton dynamics in 1L-WSe₂. The radiative recombination rates ($\gamma_X$, $\gamma_D$ and $\gamma_{X_K}$) and the inter-state conversion rates between X, D and $X_K$ excitons ($\gamma_{X-D}$, $\gamma_{X-X_K}$, and $\gamma_{2D-2X_K}$) are labelled accordingly. Right: The spin-valley configurations of the corresponding inter-state conversion. **d**, the Tr-PL of D and $X_K$ exciton under resonant excitation at pump fluence of 0.13 μJ/cm⁻² at 1.72 eV. The data is fitted with the rate equation model described in **c**. **e**, the evolution of the exciton density with initial condition $N_X = 7.5 \times 10^{10}$ cm⁻² Inset: a comparison of the contribution from X to $X_K$ and D to $X_K$ inter-state conversions.

Figure 2a displays a typical contour plot of Tr-PL spectra of our sample taken by the streak camera: the x-axis is the emission photon energy, and the y-axis is the time *t*. The time zero is approximated by measuring the arrival time of the laser reflection from the sample. We also overlay a PL spectrum taken with CCD in Figure 2a to show the integrated peak intensity. Four prominent peaks are denoted from high to low energy: X, XD, D, and $X_K$. Among them, X, D, and $X_K$ are assigned to the three exciton species as illustrated in Figure 1 and XD is the biexciton composed of X and D situated in the opposite valleys.[15] (Note that the $X_K$ peak here is actually the intervalley exciton phonon replica emission. However, we anticipate the PL intensity to be proportional to the intervalley exciton population under experimental conditions in this paper.) We note that the emission from the trion (indicated by $X^-$ at 30 meV below X) in our PL spectra is $\leq 10\%$ of that from X, suggesting that our sample is quite neutral. In Figure 2a, it can already be seen that the four emission modes behave quite differently in the time domain. For X and XD, the signal becomes strong shortly after $t = 0$ and decays quickly, indicating a short population lifetime. The D exciton emission is weak but has much longer lifetime compared with X and XD, reflecting that the ground-state D exciton has fewer decay channels. The emission from the $X_K$ exciton, however, behaves dramatically different from the others: the integrated signal is surprisingly *more* intense than the X exciton and spread out over time, suggesting both slower growth and decay rate.

In Figure 2b, we plot the evolution of PL intensity of the X, D, and $X_K$ exciton, with each normalised to its maximum intensity. After performing deconvolution with the instrument response function (IRF), we found that both X and D excitons exhibit similar double exponential decays. In contrast, the intensity of $X_K$ emission first grows slowly and reaches its maximum at about 200 ps before decreasing with a slow exponential decay. We extracted the growth and the decay lifetimes by performing a double exponential fitting of the intensity $I$ for all excitons to the function $I(t) = y_0 + A_1 \exp\left(-\frac{t}{t_1}\right) + A_2 \exp\left(-\frac{t}{t_2}\right)$. The fitting results are plotted along with the experimental data in Figure 2b and summarised in Table 1 as well.

For the X exciton, the fast decay is associated with the intrinsic radiative recombination.[5] However, because the time resolution in our setup is about 5 ps, the $t_1 = 4$ ps is subject to having large uncertainty, and likely represents an upper bound of the radiative recombination lifetime. In addition to the initial fast decay, we observe a significant contribution from the second decay component $t_2 = 215$ ps. The prominent two-time-constant behaviour has been previously attributed to the inter-state conversion between the bright X exciton and the underlying dark excitons.[28,29] In Table 1, we further estimate the fraction of total intensity emitted by this channel $Y_2 = A_2 t_2 / (A_1 t_1 + A_2 t_2)$. We found that $Y_2$ for the X exciton is > 74%, even higher than the analogous process in carbon nanotubes (CNTs; < 40%)[29], suggesting that the dark exciton in 1L-WSe$_2$ behaves as a significant reservoir of X exciton emission through efficient inter-state up-conversion.[24]

**Table 1.** The fitting results in Figure 2b. Note that the $t_1$ and $A_1$ of X$_K$ exciton are acquired from the fit of growth rate.

|       | $t_1$ (ps)  | $t_2$ (ps) | $A_1$ (a.u.)     | $A_2$ (a.u.)     | $Y_2$ |
|-------|-------------|------------|------------------|------------------|-------|
| X     | $4.0 \pm 0.1$ | $215 \pm 8$  | $0.86 \pm 0.05$    | $0.05 \pm 0.001$   | 0.74  |
| D     | $24 \pm 2.7$  | $258 \pm 11$ | $0.57 \pm 0.03$    | $0.44 \pm 0.02$    | 0.89  |
| X$_K$ | $45 \pm 1$    | $481 \pm 5$  | $-0.92 \pm 0.01$   | $1.36 \pm 0.01$    | –     |

For the D and X$_K$ excitons, we observe a much slower population decay rate, consistent with the much smaller oscillator strength compared to X. Specifically, the radiative emission of D and X$_K$ requires a spin-flip of electrons[30] and defect/phonon scattering[21–23], respectively. To capture the dynamics of all the excitons populations, especially the one of X$_K$ exciton, we perform a universal fitting with the rate equations, Eq.(1)–(3), describing the four-level (X, D, X$_K$ and the ground state) system illustrated in Figure 2c:

$$\frac{dN_X(t)}{dt} = -\left(\gamma_X + \gamma_{X-D} + \gamma_{X-X_K}\right)N_X(t) + \gamma_{X-D}N_D(t) + \gamma_{X-X_K}N_{X_K}(t) \tag{1}$$

$$\frac{dN_D(t)}{dt} = -(\gamma_D + \gamma_{X-D})N_D(t) + \gamma_{X-D}N_X(t) + \gamma_{2D-2X_K}\left(N_{X_K}(t)^2 - N_D(t)^2\right) \tag{2}$$

$$\frac{dN_{X_K}(t)}{dt} = -\left(\gamma_{X_K} + \gamma_{X-X_K}\right)N_{X_K}(t) + \gamma_{X-X_K}N_X(t) + \gamma_{2D-2X_K}\left(N_D(t)^2 - N_{X_K}(t)^2\right) \tag{3}$$

where $N_X(t)$, $N_D(t)$, and $N_{X_K}(t)$ are the densities of X, D, and X$_K$ excitons, respectively; $\gamma_X$, $\gamma_D$, and $\gamma_{X_K}$ are the corresponding radiative recombination rates; and $\gamma_{X-D}$, $\gamma_{X-X_K}$, and $\gamma_{2D-2X_K}$ are the inter-state conversion rates. In order to achieve a good description of the experiment, we need to introduce the $\gamma_{2D-2X_K}$ terms, which are associated with the second order in density. This is consistent with the efficient conversion through the Coulomb exchange interaction, as illustrated in Figure 2c. As we discuss below, the second-order dependence on density is required to fit the fluence-dependent data.

Resonant excitation of the X exciton in the low-density regime allows us to establish a well-defined initial condition of the X exciton density. We resonantly excite the sample with pump fluence of $0.13$ μJ·cm$^{-2}$, corresponding to the estimated initial X exciton density of $7.5 \times 10^{10}$ cm$^{-2}$ (see Supplementary S1 for the estimation of exciton density.) Figure 2d shows the resulting Tr-PL along with the fits to our model, for both D and $X_K$ excitons. The fits allow us to determine four parameters: $\gamma_X^{-1} = 4$ ps, $\gamma_D^{-1} = 2,000$ ps, $\gamma_{X-D}^{-1} = 150$ ps, and $\gamma_{2D-2X_K}^{-1} = 8.3 \times 10^{10}$ ps·cm$^2$. We furthermore determine that the times $\gamma_{X_K}^{-1}$ and $\gamma_{X-X_K}^{-1}$ exceed $10,000$ ps, indicating that these processes are negligible in our experiment. The negligible $\gamma_{X-X_K}$ indicates an inefficient phonon coupling, as expected at cryogenic temperatures. The other two inter-state conversion rates, $\gamma_{X-D}$ and $\gamma_{2D-2X_K}$, are quite significant. The large $\gamma_{X-D}$ suggests an efficient conversion between X and D excitons due to spin-flip assisted by the spin-orbit interaction,[30] in agreement with the prior discussion of the large $Y_2$ of the X exciton. The negligible $\gamma_{X-X_K}$ suggests that the direct process $D \leftrightarrow X_K$ should also be negligible as it requires a defect or phonon. However, the second-order process $\gamma_{2D-2X_K}$ is comparable in magnitude to the first-order process $\gamma_{X-D}$, providing a superlinear conversion between D and $X_K$ excitons as discussed below.

The significant conversion rate $\gamma_{X-D}$ and $\gamma_{2D-2X_K}$ leads to a long-lived and high-density population of D and $X_K$ excitons. In Figure 2e, we simulate the evolution of the exciton density to reveal the interplay between the three exciton species. In the condition of resonant excitation at 4.2 K, most of the X excitons are within the light cone as generated. As a result, the population of X exciton significantly drops through the radiative recombination. Only a fraction of X excitons $\gamma_X^{-1}/\gamma_{X-D}^{-1} = 4$ ps/150 ps are converted to D excitons, and then achieve the maximum population of D exciton $2 \times 10^9$ cm$^{-2}$ at $t = 12$ ps, matching our observation of the fast growth rate of D exciton in Figure 2b. For the $X_K$ exciton, as shown in the inset of Figure 2e, the rate $\gamma_{2D-2X_K}N_D^2$ is about 3 to 4 orders of magnitude larger than our upper-bound estimate of the rate $\gamma_{X-X_K}N_X$ during almost the whole time frame, suggesting that the population of $X_K$ exciton is created mainly through the upconversion from the D exciton. Our results reveal that the Coulomb exchange interaction from D and $X_K$ exciton indeed plays an essential role in the population dynamics.

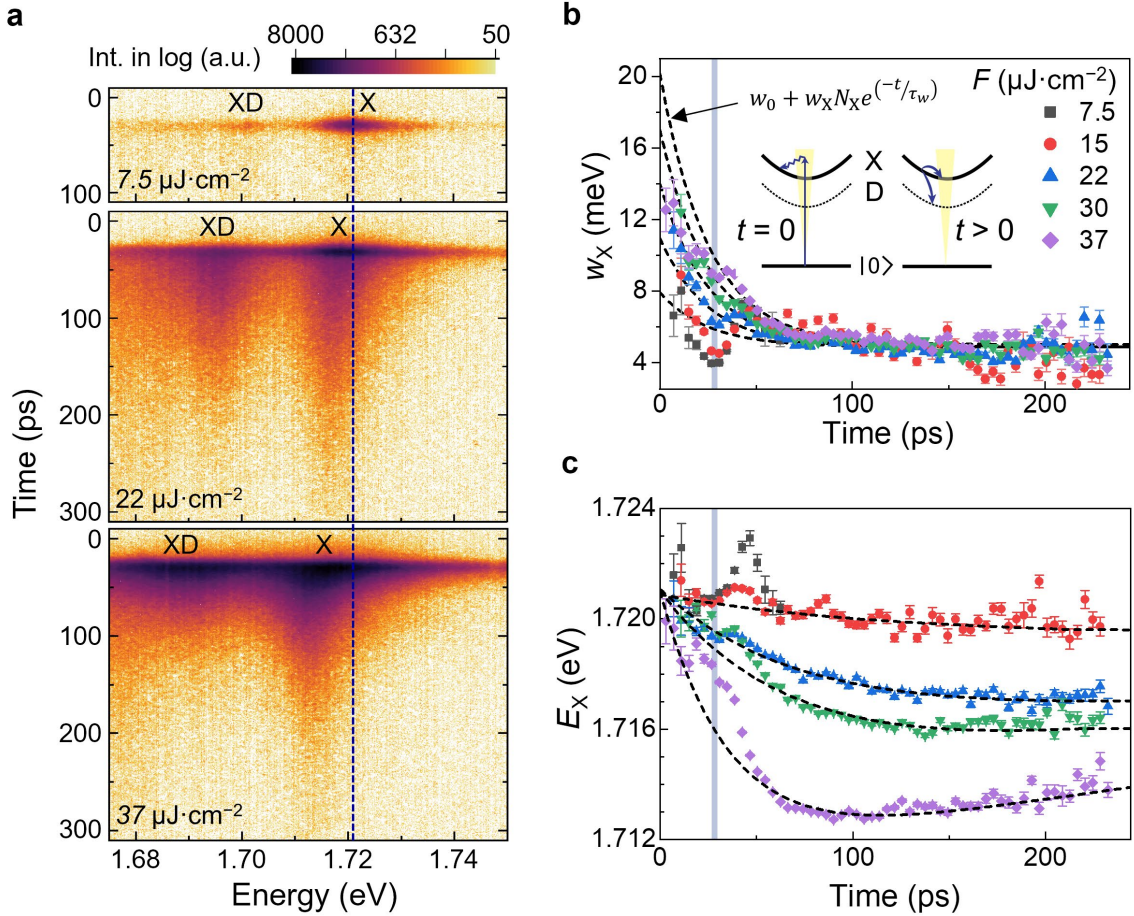

**Figure 3. Evolution of the peak energy and the linewidth of X exciton at high excitation fluence. a,** the contour plots (PL intensity in log) of Tr-PL spectra at excitation fluence of 7.5, 22 and 37 μJ·cm$^{-2}$ with near resonant excitation at 1.78 eV (60 meV above $E_X$). The vertical dashed line is aligned with the peak energy of X at the lowest fluence. **b,** the fluence-dependent linewidth of X exciton as a function of time. The dashed curves are the fitting results of exciton-exciton interaction mediated linewidth broadening. **c,** the fluence-dependent peak energy of X exciton as a function of time. The dashed curves are fitted with double exponential functions. The grey vertical lines in **b** and **c** are aligned with the timing of the maximum intensity at around 30 ps.

Taking advantage of the streak camera, we turn to studying the evolution of the spectral shape of PL emission. Here, we perform Tr-PL with a near-resonant excitation at 60 meV above the X exciton. In Figure 3a, we display the contour plots of Tr-PL spectra taken under various excitation fluence. At 7.5 μJ·cm$^{-2}$, the features are similar to Figure 2a. As the fluence increases to 37 μJ·cm$^{-2}$, we observe apparent biexciton features which grow super-linearly with the excitation density.[15] More interestingly, the time-dependent emission of X exciton shows an asymmetrical shape within the extended tail, indicating an excitation fluence-sensitive interaction between the X exciton and the dark excitonic states.

To resolve the evolution of the PL spectra, we perform a multipeak fitting with Lorentzian functions. The evolution of the linewidth and the peak energy of the X exciton are shown in Figure 3b and 3c, respectively. The origin of the asymmetric shape in Figure 3a is primarily due to a combination of linewidth broadening as well as a persistent redshift of the peak energy. We first discuss the evolution of linewidth (defined by the half-width of the half maximum of the Lorentzian function) at different excitation fluences from 7.5 to 37 μJ·cm$^{-2}$. As can be seen in Figure 3b, the dynamics of linewidth can be described by two mechanisms acting at different timescales: 1) an initial linewidth broadening followed by an exponential decay to 4 meV at $t > 250$ ps, and 2) a sharp linewidth narrowing at around $t = 30$ ps, correlated with the maximum PL intensity. The first mechanism is mainly due to exciton-exciton interaction induced homogeneous linewidth broadening.[31] Under the near-resonant excitation, the incoming photons couple to the phonons at time zero, resulting in highly-populated hot X excitons outside the light cone. These hot X excitons then thermalise either by intra-band scattering *via* exciton-exciton and exciton-phonon interaction or inter-band scattering to the D excitons, leading to the exponential decrease of the linewidth. The origin of the peak narrowing at $t \approx 30$ ps is not clear, but could be associated with the optical Stark effect at the high photon density, as it appears coincidently with a sharp blue shift (see below).[32]

We first examine whether the linewidth broadening due to heating of the lattice mediated by exciton-phonon interactions. Quantitatively, taking the excitation fluence at 37 μJ·cm$^{-2}$ as an example, we estimate that the initial exciton density is $N_X = 6.5 \times 10^{11}$ cm$^{-2}$, given that the absorption at 1.78 eV nm is about 0.5% and assuming that the conversion rate from the incident photons to excitons is unity. Under the near-resonant excitation at 60 meV above the X exciton, the excess energy of the hot X excitons is about $N_X \times 60$ meV. If we assume that the total excess energy is transferred to the lattice, the corresponding lattice temperature increase is less than 10 K (see Supplementary S2 for heat capacity calculation), leading to a nominal linewidth broadening less than 0.5 meV, which is mostly negligible.[15]

We instead consider that the linewidth broadening as the result of exciton-exciton interactions, leading to an exciton density-dependent linewidth[31] $w(N_X) = w_0 + w_X N_X(t)$, where, $w_0 = 4.5$ meV is the zero-excitation density linewidth taken from the linewidth at $t = 250$ ps where $N_X$ is negligible), $w_X$ is the coefficient relating the density of X exciton to the linewidth. By plugging in the estimated $N_X$ at time zero for all different fluences, as shown in the fitting curves in Figure 3b, we can describe the evolution of the linewidth by a single exponential decay with $w_X = 1.4 \times 10^{-11}$ meV·cm$^2$, and the time constant of $\tau_w = 25$ ps. We note that the above fittings are done by excluding the peak narrowing effect at $t \approx 30$ ps. Our results suggest that both exciton-phonon and exciton-exciton interaction of X exciton should only contribute in $t < 100$ ps, where the population of X exciton significant drops *via* both the efficient radiative recombination and the X to D inter-state conversion.

Next, we study the dynamics of the peak energy of the X exciton. As can be seen in Figure 3c, the timescales of the redshift dynamics are dramatically different from the observed linewidth dynamics. The evolution of the peak energy exhibits an initial slow redshift followed by a slow blueshift. (A narrowly timed blueshift around $t \approx 30$ ps may be related to the similarly timed linewidth narrowing discussed above.) At the lower fluence $F = 15$ μJ·cm$^{-2}$, the redshift seems to persist beyond the timeframe of our measurement. Even at the highest fluence $F = 37$ μJ·cm$^{-2}$, the magnitude of the redshift reaches a maximum at about 100 ps and then decays very slowly. This behaviour is substantially similar to the evolution of $X_K$ exciton population shown in Figure 2e.

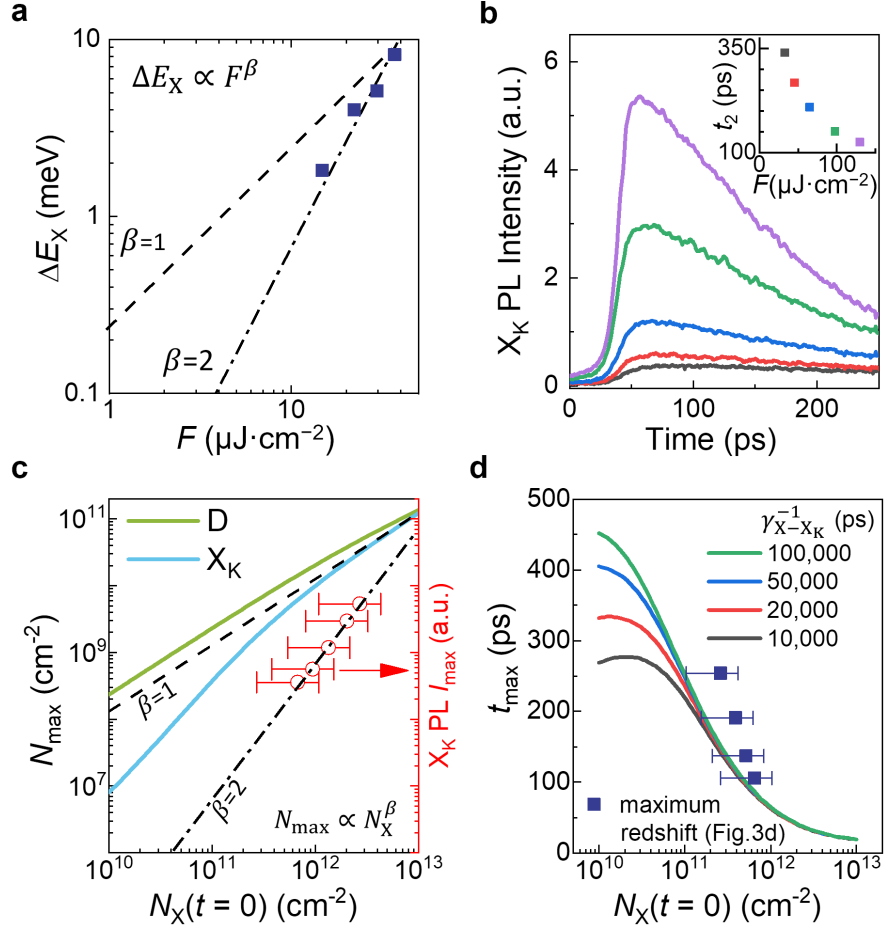

**Figure 4. The dynamics and the excitonic screening effect of the $X_K$ exciton. a,** the magnitude of redshift extracted in Figure 3c at various excitation fluence. The dashed and dot-dashed lines are corresponding to the power law of $\beta = 1$ and $\beta = 2$ to the fluence, respectively. **b,** the fluence-dependent dynamics of the $X_K$ exciton. The fluence-dependent decay time $t_2$ is plotted in the inset. **c,** left y-axis: The maximum population of the D and $X_K$ excitons as function of the initial population of the X exciton ($N_X$). Right y-axis: The maximum PL intensity of $X_K$ exciton extracted from **b**, which follows a power law of $N_X^2$. **d,** The navy solid squares are the timing of the maximum redshift extracted from Figure 3c. The solid curves are the simulation results of the timing of maximum population of $X_K$ exciton with various $\gamma_{X-X_K}$.

To be quantitative, we perform double exponential fitting to extract the magnitude of the maximum redshift and plot it against the fluence in Figure 4a. Interestingly, the redshift of the X exciton can be approximately described by a power law, $\Delta E_X \propto F^2$, reflecting that the dynamics of the redshift is quite sensitive to the excitation fluence. The superlinear fluence dependence, as well as the slow response time of the redshift, helps us to elucidate the underlying mechanisms. We first consider whether the redshift could be due to the laser heating effect. In our case, the near-resonant excitation creating the exciton density of $\sim 10^{12}$ cm$^{-2}$ should give a negligible heating effect, as the estimated lattice temperature increase of <10 K would correspond to <2 meV of bandgap narrowing[33] and <0.4 meV redshift in X exciton.[15] Furthermore, heating should be accompanied by a linewidth broadening due to the exciton-phonon interaction, however in our examination of the linewidth above, we already argued that the laser-induced heating, if any, should be fully thermalised within 100 ps, in contrast to the observed slowly evolving and persistent redshift. Lastly, heating should be linear in fluence.

An alternative explanation of the redshift is the excitonic screening effect. The excitonic Coulomb interaction can reduce both the electronic bandgap and the exciton binding energy, resulting in a net redshift of the X exciton state.[33] We can first rule out the excitonic screening from X excitons because $N_X$ drops sharply in 25 ps and keep decreasing exponentially, in contrast to the nonmonotonic redshift. However, the populations of D and $X_K$ excitons persist for a much longer time. Particularly, the $X_K$ exciton has a superlinear conversion rate from the D exciton, matching our observation that the maximum of the exciton population is superlinear in the incident fluence.

We then examine the relation of the density of the $X_K$ exciton to the excitation density. Figure 4b displays the Tr-PL of the $X_K$ exciton at various incident fluence, and the extracted maximum intensities $I_{max}$ are plotted as open red circles in Figure 4c. As can be seen, the trend can be well described by the dependence $N_{X_K-max} \propto N_X^2$. We note that the behaviour of $I_{max}$ should not be confused with the power dependent peak intensity in the continuous-wave (CW) measurement.[22,23] In Supplementary S3, we show the power-dependent PL measurements with a CW laser at 2.33 eV, where the intensity of $X_K$ exhibits a linear dependence on the incident power. This discrepancy can be understood as illustrated in the inset of Figure 4b: Although the maximum intensity is growing super-linearly, the lifetime of the exponential decay also decreases (due to more efficient conversion $X_K \rightarrow D \rightarrow X$), resulting in a linear dependence of PL intensity. The quadratic dependence $N_{X_K-\text{max}} \propto N_X^2$ confirms our assertion that the creation of $X_K$ is governed by the second-order process $\gamma_{2D-2X_K}$. It also matches the superlinear dependence of the magnitude of the redshift on fluence in Figure 4a, strongly indicating that the $X_K$ population density is responsible for the redshift. To further support our observation, we perform the fluence-dependent simulation of the D and $X_K$ exciton populations with the rate equations Eq.(1)−(3). Using the fitting parameters acquired from Figure 2d, the $N_{D-\text{max}}$ is mostly linearly proportional to $N_X$ in the whole range, as shown in Figure 4b. In contrast, the $N_{X_K-\text{max}}$ behaves super-linearly at the low-density regime and slowly evolves to the linear behaviour at the high-density regime.

With the understanding of the distinct behaviour of D and $X_K$ excitons, we can estimate the redshift based on the excitonic screening theory. We find that both the dynamics of the redshift (Figure 3c) and the dependence on fluence (Figure 4a) are qualitatively consistent with the time evolution and fluence dependence of the $X_K$ exciton density (Figure 4b). In Supplementary S4, we estimate the coefficient $a_{X_K}$ relating the redshift to the density of $X_K$ exciton by $a_{X_K} = \Delta E_X / N_{X_K} = (1.3 \pm 0.08) \times 10^{-9}$ meV · cm$^2$. Our results reveal an essential role of the $X_K$ exciton in the excitonic screening. It is of interest to note that, although the D exciton has higher population density than the $X_K$ exciton in the initial 100 ps, its contribution to the Coulomb screening is much less. This could be because the D exciton has smaller dipole moment due to its spin configuration: The same spin of electron and hole in the D exciton leads to a larger Bohr radius compared to the X and $X_K$ excitons.

Lastly, we address the issue of whether the direct single-exciton intervalley scattering process $X_K \leftrightarrow X$ is observable. So far, we have modelled our results assuming negligible inter-state conversion $\gamma_{X-X_K}$ and recombination rates $\gamma_{X_K}$, implying the weak coupling between exciton and K-phonons. In Figure 4d, we plot the corresponding time of the maximum density $t_{\text{max}}$ extracted from Figure 3c, along with the simulation results with various scattering rates from $\gamma_{X-X_K}^{-1} = \gamma_{X_K}^{-1} = 10,000$ ps to $100,000$ ps. We note that the fits performed in Figure 2d are insensitive to both $\gamma_{X-X_K}$ and $\gamma_{X_K}$ in this range of magnitude. As demonstrated in Figure 4c, all the curves qualitatively describe the observed $t_{max}$ quite well at the exciton density range in our experiments. However, at the low-density regime, $\gamma_{X-X_K}$ and $\gamma_{X_K}$ become dominant due to the superlinear dependence of the $\gamma_{2D-2X_K}$ on density. This indicates that the direct rates $\gamma_{X-X_K}$ and $\gamma_{X_K}$ should be measurable experimentally in lower density experiments with sufficient signal-to-noise ratio.

**Conclusion**

In conclusion, through Tr-PL studies of the charge-neutral exciton complexes in 1L-WSe$_2$, we observe long-lived populations of the spin-0 intervalley X$_K$ exciton and the spin-1 intravalley D exciton, closely linked to the double exponential decay observed in the bright X exciton. Assisted by the rate equations analysis, our results provide comprehensive evidence for the efficient and superlinear conversion between two D and two X$_K$ excitons through mutual Coulomb exchange interactions. Furthermore, we demonstrate an appreciable excitonic screening effect from the highly populated X$_K$ excitons, which could enable studies of the rich many-body correlated physics for the dark excitons. Our work paves the way to study new phenomena in long-lived indirect excitons in TMDs, including many-body interactions, transport and manipulation of valley and spin, and dark excitonic condensates.

**Methods:**

**Sample Fabrication.** The bulk WSe$_2$ crystals are grown by chemical vapour transport method.[34] The 1L-WSe$_2$ and hBN flakes are first exfoliated on precleaned 300 nm SiO$_2$/Si wafers. To assure the sample quality, we employ the typical optical microscope and atomic force microscopy to select the residue-free 1L-WSe$_2$ and few-layer hBN flakes. The hBN/1L-WSe$_2$/hBN samples are then stacked using a dry transfer technique with PPC (poly-propylene carbonate) stamps using a home-made micro-alignment system. All the sample preparation processes are completed in a nitrogen-purged dry glovebox to avoid sample degradation. After stacking, the sample is further thermally annealed at 350 °C for 1 hour in argon atmosphere (99.999%) to improve the sample quality.

**Time-resolved photoluminescence measurement.** We mounted the sample in a continuous-flow cryostat (Janis ST-500) and cooled down with liquid helium at the base temperature of 4.2 K for spectroscopy measurements. The sample is excited with a linear polarised femtosecond pulsed laser (Coherent: Chameleon Ultra II) with a nominal pulse width of 140 fs. The incident laser is first spectrally cleaned by the short-pass filters (Thorlabs FESH0700 or FESH0750) and then focused by a 50× objective lens (numerical aperture=0.5). The PL signal is collected in the back-scattered geometry. To avoid the local heating effect, instead of focusing the laser in a diffraction-limited spot, we added a thin lens in the excitation path to expand the spot size up to 40 µm in diameter. The pulse-width on the sample is estimated at 250−300 ps due to the dispersion of the optics. In the collection path, we set a pinhole with a 4f optical system to perform the confocal spectroscopy, enabling us to detect the signal in the selected area with a diameter of 3 µm. The collected signal is then spectrally filtered (Thorlabs FELH0700 or FELH0750) to remove the Rayleigh scattered signal. To measure the Tr-PL with a streak camera, we first dispersed the PL signal by a monochromator (Andor Kymera 328i) with a 600 groove/mm grating, and then coupled to the streak camera (Optronis OptoScope SC-10). The time resolution in all the data shown in this work is about 5 ps, which is determined by the FWHM of IRF function. Alternatively, we could measure the PL spectra by a high-resolution spectrometer (1200 grooves/mm grating) equipped with thermal-electrically cooled charge-coupled devices. (Andor iXon EMCCD).

**Acknowledgements**

This work was supported by the Australian Research Council (ARC) through the Centre of Excellence Grant CE170100039 (FLEET). JY is supported by NSF DMR-2004474. KW and TT acknowledge support from the Elemental Strategy Initiative conducted by the MEXT, Japan, Grant Number JPMXP0112101001, JSPSKAKENHI Grant Numbers JP20H00354, and the CREST (Grant JPMJCR15F3), JST.

## S1. Estimation of the Exciton Density

In the Figure 2c−e in the main text, we employed a four-level rate equation model to fit the time-resolved photoluminescence (Tr-PL) data. It is critical to determine the initial condition in our model through estimating how many excitons are generated at time zero. Given an incident fluence $F$, we can calculate the exciton density by Eq.(S1) as follows:

$$N_X = \frac{F}{E_{\text{photon}}} \times A \qquad (S1)$$

Where $E_{\text{photon}}$ is the photon energy in Joule, $A$ is the wavelength-dependent absorptance of the 1L-WSe$_2$. This estimation gives the upper bound of the exciton density by assuming that one absorbed photon generates one exciton. A widely employed method to measure the absorptance of the 2D materials is by measuring the differential reflectance ($dR/R$) in a back-scattered geometry.[1,2] Figure S1a shows the $dR/R$ spectra of our hBN encapsulated 1L-WSe$_2$ sample at 4.2 K. As can be seen, due to the interference of the reflected light from the multiple interfaces, the $dR/R$ spectrum is highly distorted. It is therefore difficult to directly estimate the absorptance by this method. Alternatively, we found a report of the measurements for both reflectance and transmittance on a chemical vapor deposition (CVD) grown 1L-WSe$_2$ sample.[3] From the Figure 2 in the Ref. 3, we can derive the absorptance spectra, as plotted in Figure S2. We note that the sample quality of our hBN-encapsulated 1L-WSe$_2$ is better. Therefore, our sample should have larger oscillator strength and narrower linewidth. In this work, we estimate the exciton density by employing the following values: For the resonant excitation (1.72 eV), we take $A = 0.2 \pm 0.1$. For near-resonant excitation (1.78 eV), we take $A = 0.005 \pm 0.003$. The uncertainty in $A$ is represented as the error bar in x-axis of $N_X$ in the corresponding figures.

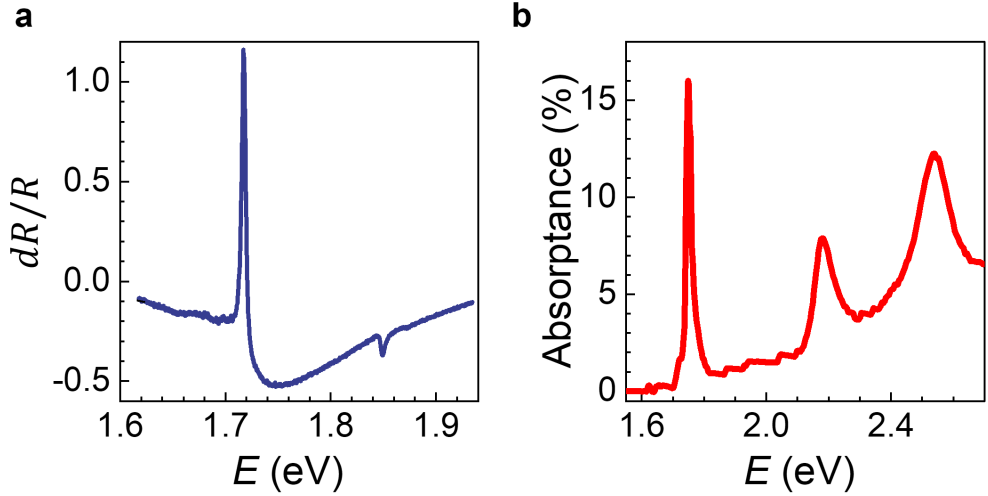

**Figure S1. a.** The $dR/R$ spectra of our hBN-encapsulated 1L-WSe₂ sample at 4 K. **b.** The absorptance of CVD 1L-WSe₂ at 10 K derived from Ref 3.

## S2. Heat Capacity of 1L-WSe₂ and the Estimation of Laser Heating Effect

In Figure 3, we applied a near-resonant excitation at 60 meV to populate the X exciton with the estimated exciton density around $N_X = 10^{12}$ cm⁻². In this section, we estimate the laser-induced heating effect by considering the heat capacity at cryogenic temperatures. Here, we first assume that every photon-excited hot exciton carries the excess energy of 60 meV. The whole excitonic system thus gains the total excess energy density $\Delta Q = N_X \Delta E = 60\text{meV} \times N_X$, yielding the exciton temperature around 700 K. Next, we consider the thermal energy from the hot excitons is fully thermalised to the lattice through exciton-phonon scattering. The increase of lattice temperature can be estimated by calculating the lattice heat capacity $C_l(T) = C_o(T) + C_a(T)$, where $C_o(T)$ is the heat capacity contributed by six optical phonon branches, and $C_a(T)$ is by three acoustic phonon branches. To simplify the calculation, we take the average phonon energy $\hbar\omega_o = 29$ meV from 6 optical phonon branches[4] ($2E'$, $2E''$, $A_1'$, $A_2''$) and calculate the contribution to $C_l$ per unit cell by:

$$C_o(T) = \frac{\partial}{\partial T}\left(\frac{6\hbar\omega_o}{e^{\hbar\omega_o/kT}-1}\right) \tag{S2}$$

For estimating the contribution from the acoustic branches per unit cell, we apply the 2D Debye model:

$$C_a(T) = 6k_b\left(\frac{T}{\Theta}\right)^2 \int_0^{\Theta/T} \frac{x^3 e^x}{(e^x-1)^2} \tag{S3}$$

where $\Theta$ is the Debye temperature $\Theta = (\hbar v / k_{\mathrm{b}})\sqrt{4\pi/A_{\mathrm{Cell}}}$. By plugging the average of the group velocity for 3 acoustic phonon branches $v \sim 10^5$ cm/s and the area per unit cell $A_{\mathrm{Cell}} = 9.1 \times 10^{-16}\,\mathrm{cm}^{-2}$ given the lattice constant $a_0 = 3.25$ Å.[5] We found that $\Theta = 84$ K in 1L-WSe$_2$. In our experimental condition, the sample temperature is 4 K, which is much lower than 84 K. Therefore, we can approximate $C_{\mathrm{a}}(T)$ in the analytic form at the low-temperature limit:

$$C_a(T) \approx 5.28 \times 10^{-4}\ T^2\ \mathrm{meV \cdot cm^{-2} \cdot K^{-1}} \qquad (S4)$$

In Figure S2a, we plot the lattice heat capacity as a function of temperature. As can be seen, the lattice heat capacity is mainly contributed by the acoustic phonon. Now we estimate the deviation of lattice temperature at various exciton density by $60\mathrm{meV} \times N_X \times A_{Cell} = \int C_l(T)dT$. We plot the three curves corresponding to different initial lattice temperature $T_i$ at 4.2 K, 10 K, and 20 K in Figure S2b. The shady area indicates the range of exciton density we estimated in Figure 3. We found even at the highest excitation fluence throughout the experiments ($N_X = 4 \times 10^{12}$ cm$^{-2}$), the lattice temperature can only increase 7 K, 3 K, and 1 K at $T_i = 4.2$ K, 10 K, and 20 K, respectively. Our results suggest that the laser heating induced linewidth broadening in our experimental condition is negligible.

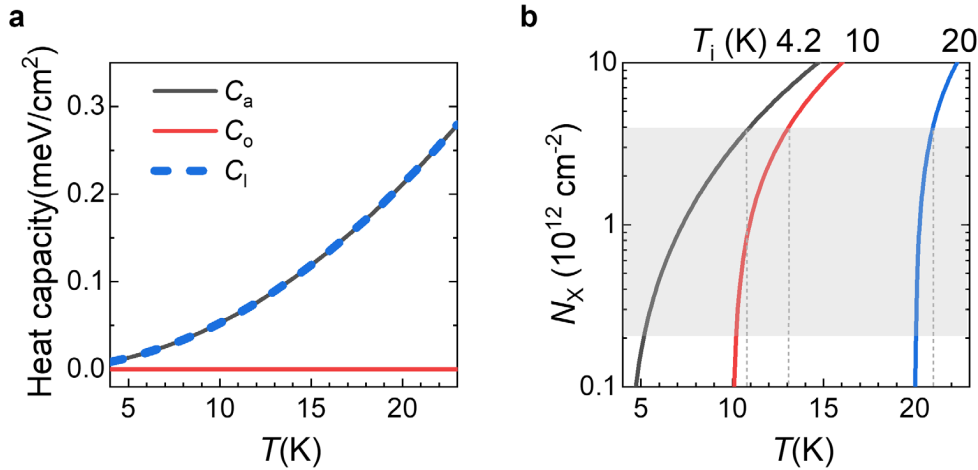

**Figure S2. a.** The temperature-dependent lattice heat capacity of 1L-WSe$_2$. **b.** The estimated increase of lattice temperature at different initial temperatures.

## S3. The Power Dependence of the $X_K$ Exciton PL Emission with Continuous-wave Excitation

In the main text, we have demonstrated that the maximum population of the $X_K$ exciton, which is superlinear to the excitation fluence. In this section, we show that the PL intensity of $X_K$ exciton is, however, linear in the excitation power under the continuous-wave (CW) excitation. As can be seen in Figure S3a, the peak intensity of the $X_K$ exciton can be extracted through the multi-peak fitting with Lorentzian functions; a selected fitting result is shown in Figure S3a. Figure S3b shows that PL intensity $I$ of the $X_K$ exciton exhibits a linear relationship to the incident power.

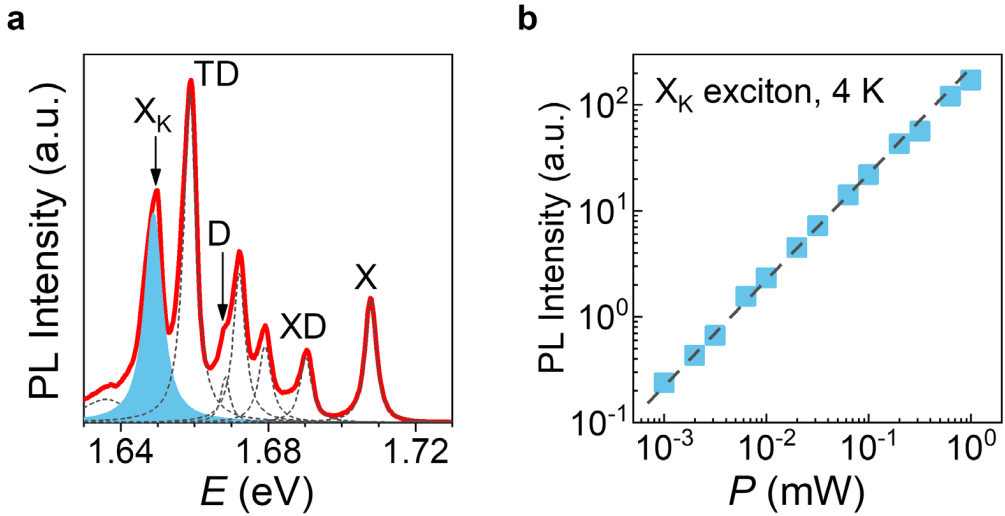

**Figure S3. a.** A selected PL spectrum of hBN-encapsulated 1L-WSe$_2$ at 4 K. **b.** The power-dependent PL intensity of $X_K$ exciton with CW laser excitation at 2.33 eV. The dashed line shows a power law, $I \propto P$.

## S4. The Comparison of the Excitonic Screening Mediated by Different Types of Exciton

In this section, we would like to estimate the relation of the population density of $X_K$ exciton to the redshift of X exciton. In Figure S4, the navy squares represent the magnitude of the redshift $\Delta E_X$ shown in Figure 4a in the main text. Here, we plot this data against the density of the maximum population of $X_K$ exciton ($N_{X_K}$). which is estimated by the simulation results (the $X_K$ curve in Figure 4c). Assuming the linear dependence at the low-density regime, we can perform a linear fitting with our data and extract the coefficient $a_{X_K}$ relating the redshift to the density of $X_K$ exciton by $a_{X_K} = \Delta E_X / N_{X_K} \approx (1.3 \pm 0.08) \times 10^{-9}$ meV $\cdot$ cm$^2$. For comparison, we also extract the data from previous studies on the excitonic screening mediated by X exciton in 1L-MoS$_2$[6] and 1L-WS$_2$.[7] We find that $a_{X_K}$ extracted in our experiment is about an order of magnitude larger than the extracted coefficient of X exciton in WS$_2$,[6] suggesting the $X_K$ excitons in our sample exhibit stronger excitonic screening effect.

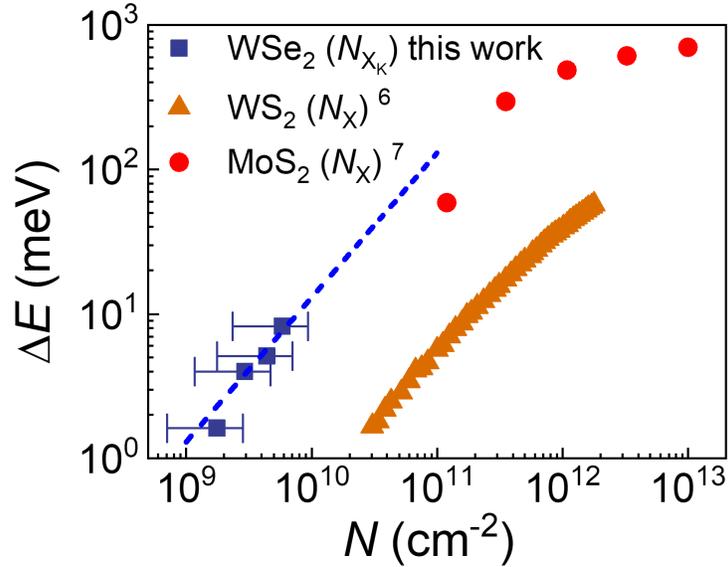

**Figure S4.** The relation of the redshift of the X exciton to the population density of the corresponding type of exciton defined in the legend. The navy squares are from our work. The blue dashed line shows the linear fitting of our data. The orange triangles and the red circles are extracted from the previous studies[6,7] of the excitonic screening mediated by the X exciton in different materials.